\documentclass[12pt]{article}
\usepackage{amsmath,amssymb}

\usepackage{graphicx}

\usepackage{hyperref}

\def\beq{\begin{eqnarray}}
\def\eeq{\end{eqnarray}}

\def\hD{\hat{\cal D}}
\newcommand{\be}{\begin{equation}}
\newcommand{\ee}{\end{equation}}
\newcommand{\bea}{\begin{eqnarray}}
\newcommand{\eea}{\end{eqnarray}}
\newcommand{\bg}{\begin{gather}}
\newcommand{\eg}{\end{gather}}
\newcommand{\bseq}{\begin{subequations}}
\newcommand{\eseq}{\end{subequations}}

\renewcommand{\ln}{\mathop{\rm ln}\nolimits}

\def\tr{\hbox{Tr}}

\def\be{\begin{eqnarray}}
\def\ee{\end{eqnarray}}
\def\lb{\label}

\begin{document}

\title{\textbf{Distributional Geometry of Squashed Cones}}

\vspace{1cm}
\author{Dmitri V. Fursaev$^{\star,\flat}$, Alexander Patrushev$^{\flat}$ and Sergey N. Solodukhin$^{\sharp}$} %\copyright

\date{}
\maketitle

\begin{center}
\hspace{-0mm}
  \emph{$^{\star}$ Dubna International University }\\
   \emph{  Universitetskaya str. 19, 141 980, Dubna,}\\
     \emph{ Moscow Region, Russia}
\end{center}     
\begin{center}
\hspace{-0mm}
\emph{$^{\flat}$ The Bogoliubov Laboratory of Theoretical Physics}\\
 \emph{ Joint Institute for Nuclear Research}\\
  \emph{Dubna, Russia\\}
\end{center}
\begin{center}
  \hspace{-0mm}
  \emph{ $^{\sharp}$ Laboratoire de Math\'ematiques et Physique Th\'eorique  CNRS-UMR
7350 }\\
  \emph{F\'ed\'eration Denis Poisson, Universit\'e Fran\c cois-Rabelais Tours,  }\\
  \emph{Parc de Grandmont, 37200 Tours, France} \\
  \emph{and  	
Institut des Hautes Etudes Scientifiques (IHES), 35 rue de Chartres, }\\
   \emph{91440 Bures-sur-Yvette, France}\\
\end{center}
{\vspace{-14cm}
\begin{flushright}
 IHES/P/13/21
\end{flushright}
\vspace{12.5 cm}
}

%\hfill{\tt IUB-TH/***}\\\mbox{} \\
%@twocolumnfalse\endcsname

%\maketitle \thispagestyle{empty}% \vspace*{.5cm}

\begin{abstract}
\noindent {A regularization procedure developed in \cite{Fursaev:1995ef} for the
integral curvature invariants on manifolds with conical singularities 
is generalized to the case of squashed cones. In general, the squashed conical singularities 
do not have rotational $O(2)$ symmetry in a subspace orthogonal to a singular surface
$\Sigma$ so that the surface  is allowed to have extrinsic curvatures. 
A new feature of the squashed conical
singularities is that  the surface terms in  the integral invariants, in the limit of small  angle deficit, now depend also on the extrinsic curvatures of $\Sigma$.
A case of invariants which are quadratic polynomials
of the Riemann curvature is elaborated in different dimensions and applied to several problems related 
to entanglement entropy. 
The results are in complete agreement with 
computations of the logarithmic terms in entanglement entropy of 4D conformal theories \cite{Solodukhin:2008dh}. Among other applications of the suggested method are
logarithmic terms in entanglement entropy of non-conformal theories and a holographic formula
for entanglement entropy in theories with gravity duals.
}
%\noindent {PACS: 04.70Dy, 04.60.Kz, 11.25.Hf }}
\end{abstract}
%\vskip 2.pc
%\maketitle

\vskip 2 cm
%\noindent $^{\star}$  e-mail: Sergey.Solodukhin@lmpt.univ-tours.fr

\newpage
    \tableofcontents
\pagebreak

\newpage

\section{Introduction and main results}
\setcounter{equation}0

The fact that the curvature of manifolds with conical singularities has a distributional nature is well known since the work by Sokolov and Starobinsky in 1977
\cite{Sokolov-Starobinsky} who studied a spacetime around a straight cosmic string. Later, 
this result was extended to certain invariant
curvature polynomials such as the Euler number and the Lovelock gravity \cite{Banados:1993qp}. 
A completely general description of delta-function terms in the curvature 
has been developed in \cite{Fursaev:1995ef}. The idea here
was to employ a regularization procedure which treats a conical space as limit of a sequence of regular manifolds.

Let $2\pi\alpha$ be a length of a unit radius circle around a tip 
of a conical space ${\cal C}_\alpha$. Near a conical singularity a manifold $\cal M$ has locally the structure of a direct product
${\cal C}_\alpha \times \Sigma$. We call codimension 2 hypersurface $\Sigma$ a singular surface. As was 
shown in \cite{Fursaev:1995ef} for integrals of some power of curvature 
tensor, if $\alpha$ is close to 1,
the leading terms proportional to $(1-\alpha)$ do not depend on 
the regularization procedure
while terms proportional to $(1-\alpha)^k$, $k\geq 2$, depend on the regularization
and are ill defined. The analysis 
of \cite{Fursaev:1995ef} 
was restricted by an assumption about a Killing vector field  
(and a corresponding $O(2)$ isometry) for which the conical singularities are fixed points. 
This condition implies that $\Sigma$ is embedded in $\cal M$ with
vanishing extrinsic curvatures. The assumption about the Killing
field was related to applications  of the conical singularity method
to (classical and quantum) entropy of black holes, see  \cite{reviews}. 
In this case $\Sigma$ is an analog of the bifurcation surface 
of Killing horizons of a stationary black hole which is a set of fixed points for an Abelian 
group of isometry.

In the more general cases, there is no $O(2)$ isometry group for which 
the singular surface $\Sigma$ is a fixed point set (although, as we will see, 
$\Sigma$ is a fixed set for a discrete group and in some
cases $\Sigma$ may be also interpreted as a bifurcation surface for event past and future horizons). 
In computations of entanglement entropy in field theories 
by a replica method $\Sigma$  has a meaning of an arbitrary entangling surface
embedded, in a simplest case, in a flat spacetime. 
Indeed, choosing the appropriate polar-like coordinates near this surface
the Minkowski spacetime metric can be written in the form
\be
ds^2=dr^2+r^2d\tau^2+(a+r\cos\tau)^2d\varphi^2+dz^2\, ,
\lb{1}
\ee
if $\Sigma$ is a cylinder of radius $a$, and
\be
ds^2=dr^2+r^2d\tau^2+(a+r\cos\tau)^2(d\theta^2+\sin^2\theta d\varphi^2)\, ,
\lb{2}
\ee
if $\Sigma$ is a sphere of radius $a$. In the both cases
the position of $\Sigma$ is at $r=0$. Suppose that $\tau$ in 
(\ref{1}), (\ref{2}) ranges from 0 to $2\pi n$, where $n=2,3,..$. A physical motivation for 
this step is described in Sec. \ref{motiv}. As a mathematical consequence
one can easily see that the geometry acquires conical singularities 
located at $r=0$ with a usual structure ${\cal C}_n \times \Sigma$. 
However, here this structure is local and is not extended to a global product of two spaces.
An important feature which should be emphasized is that
instead of a continuous $O(2)$ symmetry of a conical space 
one has a discrete group of transformations $\tau\to \tau + 2\pi k$. 
We call conical singularities of this type {\it squashed cones} by following
a terminology first used by J. Dowker in  \cite{Dowker:1994bj}.

The aim of this paper is to develop a method of calculating the integrals 
of polynomial curvature invariants in the case of the squashed conical singularities. 
We do it by purely geometric methods similar (but not identical) to those used in \cite{Fursaev:1995ef}. First of all, we test the new method for the quadratic polynomials
in different dimensions. This allows us to set the stage for future extensions.

To summarize our main results we consider a set of orbifold constructions ${\cal M}_n$
made by gluing $n$ identical replicas obtained by cutting a Riemannian manifold $\cal M$. 
The cuts are made along a codimension one hypersurface in $\cal M$ 
which ends on a codimension two hypersurface $\Sigma$. When $n$ replicas are glued together
in ${\cal M}_n$ one gets conical singularities on $\Sigma$ with the local structure
${\cal C}_n \times \Sigma$. We suppose that in integrals on a sequence of regularized
manifolds $\tilde{\cal M}_n$
there is a way to go to continuous values of $n$ and consider the limit $n\to 1$.
 
In this limit we first extend application of the known integral formula  
\be
\int_{\tilde{\cal M}_n} \sqrt{g}d^dx~ R \to n\int_{{\cal M}} \sqrt{g}d^dx~ R +4\pi(1-n)A(\Sigma)+..\, ,
\lb{0}
\ee
where $A(\Sigma)$ is the area of $\Sigma$. The difference with $O(2)$ symmetric conical 
singularities is in the appearance of the regularization dependent $O((1-n)^2)$ terms
denoted by dotes in the right hand side (r.h.s.) of (\ref{0}). In 
two dimensions, $d=2$, the  $O((1-n)^2)$ terms are absent and the formula is exact.

To present our results we define the two extrinsic 
curvature tensors of $\Sigma$
$$
k^{(i)}_{\mu\nu}=h_\mu^\lambda h_\nu^\rho (n_i)_{\lambda;\rho}~~,~~h_\mu^\lambda=\delta_\mu^\lambda-\sum_i(n_i)_\mu (n_i)^\lambda\, ,
$$
where $n_i$, $i=1,\, 2$ are two unit mutually orthogonal
normal vectors to $\Sigma$. It is convenient to introduce the following
two invariants:   
$$
\tr~ k^2=\sum_i(k^{(i)})_{\mu\nu}(k^{(i)})^{\mu\nu}~~,~~k^2=\sum_i(\tr~ k^{(i)})^2\, ,
$$ 
which do not depend on the particular choice of the pair of normals $n_i$.
Then in the limit $n\to 1$ we have the following behavior of the regularized integrals:
\be
\int_{\tilde{\cal M}_n}\sqrt{g}d^dx~R^2  \to
n\int_{{\cal M}}\sqrt{g}d^dx~R^2 +
8\pi(1-n)\int_\Sigma \sqrt{\gamma}d^{d-2}y~R+..\, , 
\lb{5.1}
\ee
\be
\int_{\tilde{\cal M}_n}\sqrt{g}d^dx~R_{\mu\nu}^2\to n\int_{{\cal M}}
\sqrt{g}d^dx~R_{\mu\nu}^2 +4\pi(1-n)\int_\Sigma \sqrt{\gamma} d^{d-2}y~\left(R_{ii}-\frac{1}{2} k^2\right)+..\, ,
\lb{5.2}
\ee
\be
\int_{\tilde{\cal M}_n}\sqrt{g}d^dx~R_{\mu\nu\alpha\beta}^2
\to n\int_{{\cal M}}\sqrt{g}d^dx~R_{\mu\nu\alpha\beta}^2
+8\pi(1-n)\int_\Sigma \sqrt{\gamma} d^{d-2}y~(R_{ijij}-\tr~ k^2)+..\, .
\lb{5.3}
\ee
The dotes in the r.h.s. of (\ref{5.1})-(\ref{5.3}) represent 
regularization dependent $O((1-n)^2)$ terms. 
Quantities $R_{ijij}$ and $R_{ii}$ have been introduced in \cite{Fursaev:1995ef}
and are invariant projections  on a subspace orthogonal to $\Sigma$
of the corresponding components of the Riemann tensor. 
If $\Sigma$ has zero extrinsic curvatures then  eqs. (\ref{5.1})-(\ref{5.3})
reduce to formulas obtained earlier in \cite{Fursaev:1995ef}.

One of main applications of our results is the calculation of entanglement entropy, see \cite{reviews} for reviews.
In particular, the integrals of the Euler density and of the Weyl squared appear in the logarithmic terms in the entropy.
The corresponding contributions on $\Sigma$ and their dependence on the extrinsic curvature have been found in \cite{Solodukhin:2008dh}
using the conformal symmetry and the  holography. It was one of motivations for the present work  to derive these surface terms
in a purely geometric way thus stressing their uniqueness and  universality.

The rest of the paper is organized as follows. In Section \ref{motiv} we
describe the physical motivations for orbifold geometries (\ref{1}), 
(\ref{2}) since these two examples play a crucial role in the subsequent analysis.
The main idea is that the cylindrical and spherical singular surfaces may be viewed
as entangling surfaces and as the bifurcation surfaces of the past and future horizons 
for sets of the specially chosen Rindler observers (although the horizons in question are not  the Killing horizons).

In Section 3 we describe our regularization method and present the calculation of integrals 
(\ref{5.1})-(\ref{5.3})
over squashed cones, first for spaces (\ref{1}), 
(\ref{2}) and then for a general geometry.  For dimensional reasons, there are only two terms which are quadratic in the extrinsic curvature, $\tr k^2$ and $k^2$, that may contribute to the integral over $\Sigma$.
Therefore, in order to fix the unknown coefficients at those terms 
it is enough to do the calculations for two typical cases (\ref{1}) and (\ref{2}).
An important feature here is that a naive application of the regularization used in \cite{Fursaev:1995ef} produces a metric which is 
singular at $r=0$.
This can be cured by introduction of an extra regularization parameter depending 
on $n$. We choose a simplest option and replace the factors $(a+r\cos\tau)$ in (\ref{1}), 
(\ref{2}) with $(a+r^nc^{1-n}\cos\tau)$, where $c$ is an irrelevant constant.
In this Section we also check the consistency of (\ref{5.1})-(\ref{5.3}) by calculating
in $d=4$  the integrals of the Euler density and of the square of the Weyl tensor.

Section 4 demonstrates a number of immediate applications of these results. It starts
with a discussion of properties of the coefficients in the heat kernel  asymptotics
of the Laplace type operators on  manifolds with the squashed conical singularities.
We then discuss the calculation of entanglement 
entropy by a replica method. We do this in terms of an effective action  
in the presence of the squashed cones, and offer some general predictions for the non-conformal field theories. 
We observe that there is no any extrinsic curvature contribution to the integral of the square of the Ricci scalar, see (\ref{5.1}). This is the only 
possibility
in the context of the entanglement entropy calculation. It indicates that the logarithmic term in the entropy is the same for a minimal and for a 
conformally coupled
scalar field. Indeed, the non-minimal coupling in the scalar field operator, $-\Delta+\xi R$, is irrelevant as soon as the spacetime is flat or Ricci flat so that, in particular, 
 it should not appear
in the logarithmic term in the entropy for these spaces.
This also agrees with the direct numerical evaluations \cite{numerical} of the logarithmic term available in the literature, which shows that the logarithmic 
term is indeed
the same for  the conformally coupled scalar field and for a minimal scalar field.

In subsection \ref{HF}  we discuss a holographic formula for entanglement entropies in conformal
field theories which allow a dual description in terms of AdS gravity one dimension higher.
We make a proposal for the holographic formula when the gravity theory includes
arbitrary terms quadratic in curvatures. This proposal extends the holographic formula
beyond the Gauss-Bonnet AdS gravities discussed earlier. In section 4.6 we define a classical gravitational entropy 
for surfaces which are not Killing horizons. This entropy is a generalization of the
Bekenstein-Hawking entropy to the case when the spacetime is not necessarily static but admits a discrete isometry.

Conclusions  are presented in Section 5. In the Appendix we present calculations of  regularized integrals in 5 and 6 dimensions.

\section{Motivations and a geometrical setup}\label{motiv}
\setcounter{equation}0

\subsection{Cylindrical and spherical Rindler horizons}\label{RH}

One physical application of distributional properties of the Riemannian geometry
in the presence of squashed conical singularities is related to the study of the entanglement
of quantum correlations across surfaces of different shapes. Such surfaces are called
entangling surfaces. In a framework of our discussion entangling surfaces
correspond to singular surfaces, so we denote these surfaces by the same letter $\Sigma$. To  make a direct link with geometries (\ref{1}), (\ref{2}),
we begin with a particular case of entangling surfaces in Minkowski
spacetime,
\begin{equation}\label{s1.3}
ds^2=-dt^2+dx^2+dy^2+dz^2\, ,
\end{equation}
which can be related to properties of event horizons for certain classes
of Rindler observers. A Rindler observer moves in (\ref{s1.3}) 
along one of the axis with a constant acceleration vector square $w^2=w_\mu w^\mu$.
As is known, the Rindler observer perceives the Minkowski vacuum as a thermal bath with 
the Unruh temperature
$T_U=w/(2\pi)$. These thermal properties result from an information loss behind 
the Rindler horizons.
The trajectory of a Rindler observer moving along the $x$-coordinate is
\begin{equation}\label{s1.4}
x(\lambda)=w^{-1} \cosh {\lambda w}~~,~~t(\lambda)=w^{-1} \sinh {\lambda w}\, ,
\end{equation}
where $\lambda$ is a proper time and the acceleration is $w$.

A well known set of Rindler observers
are those which {\it all} move in the same direction and make the so called Rindler frame
of reference
\begin{equation}\label{s1.5}
ds^2=-r^2 d\tau^2+dr^2+dy^2+dz^2\, .
\end{equation}
The transition from (\ref{s1.3}) to coordinates (\ref{s1.5}) is motivated by (\ref{s1.4})
and
has the form
\begin{equation}\label{s1.6}
x=r \cosh \tau~~,~~t=r \sinh \tau\, .
\end{equation}
All observers which are at rest with respect to (\ref{s1.5}) are the Rindler observers.
The future and past event horizons  are null hyperplanes 
which intersect (or bifurcate) at a codimension 2 plane $\Sigma$ with coordinates $x=0$,
$t=0$. $\Sigma$ divides the
states of the theory defined on a constant time hypersurface $\cal H$ ($t=0$) on those
which are located on the same side of the horizon (${\cal H}_R$, $x>0$) and can be measured, and unobservable
states on the opposite side (${\cal H}_L$, $x<0$).
One denotes these states with letters '$R$' and '$L$',
respectively. The trace of the Minkowski vacuum over the left (unobservable states)
yields the reduced density matrix of the Rindler observers
$\hat{\rho}_R=\mbox{Tr}_L~|0 \rangle\langle 0|$.

A replica method to study the quantum entanglement across $\Sigma$
is based on calculation of quantities $\mbox{Tr}_R\hat{\rho}_R^n$ where $n$ is a natural
number $n$. By ignoring the technicalities we give a geometrical construction
associated with $\mbox{Tr}\hat{\rho}_R^n$.
In a path integral representation of the Minkowski vacuum $|0 \rangle$
field configurations are set
on a half of the Euclidean space
\begin{equation}\label{s1.7}
ds^2=dt^2+dx^2+dy^2+dz^2\, ,
\end{equation}
below the hypersurface $t=0$ (i.e. for the values $t<0$).
The density matrix $\hat{\rho}_R$ is obtained
by gluing two identical half planes along their ${\cal H}_L$ parts. This yields a plane with a cut along ${\cal H} _R$. The arguments $\phi_+$, $\phi_-$ in 
the matrix elements
$\langle  \phi_+ |\hat{\rho}_R|\phi_- \rangle$ are defined on  the `upper' and `lower' parts
of the cut, which we denote as ${\cal H}_R^+$ and ${\cal H}_R^-$, respectively.
The space for the product of $n$ matrices $\hat{\rho}_R^n$ is obtained from $n$ replicas
by gluing ${\cal H}_R^-$ cut of $k$-th replica with ${\cal H}_R^+$ cut of $k+1$-th replica.
To get the trace $\mbox{Tr}\hat{\rho}_R^n$ one glues together the remaining open ends.
The corresponding space is nothing  but a higher dimensional
generalization of a Riemann surface which can easily be defined in the Euclidean
Rindler coordinates
\begin{equation}\label{s1.8}
ds^2=r^2 d\tau^2+dr^2+dy^2+dz^2\, ,
\end{equation}
where $\tau$ varies from 0 to $2\pi n$.
We denote this space ${\cal M}_n$. There is a conical singularity at
the Euclidean horizon $r=0$.
More rigorous arguments
can be given which show that $\mbox{Tr}\rho_R^n$ has a path integral representation where field configurations live on ${\cal M}_n$.

It is interesting to note that the planar Rindler horizon associated
to the Rindler coordinates (\ref{s1.5}) allows us to make a straightforward generalization to curved horizon (entangling) surfaces in Minkowski spacetime.
Consider, for example, a set of Rindler observers which move
radially with respect to the  axis $x=y=0$ and have trajectories
\be\label{s1.9}
&&x(\lambda)=a+w^{-1} \cosh \left({\lambda w}\right)\cos\varphi ~~,~~
y(\lambda)=a+w^{-1} \cosh \left({\lambda w}\right)\sin\varphi \, ,\nonumber \\
&&t(\lambda)=w^{-1} \sinh {\lambda w}~~,~~z=\mbox{const}\, ,
\ee
where $a$ is constant and $w$ is an acceleration. The corresponding
coordinate transformations
\begin{equation}\label{s1.10}
x=a+r \cosh \tau\cos\varphi~~,~~
y=a+r \cosh \tau\sin\varphi~~,~~
t=r \sinh \tau~~
\end{equation}
change (\ref{s1.3}) to
\begin{equation}\label{s1.11}
ds^2=-r^2 d\tau^2+dr^2+(a+r \cosh \tau)^2d\varphi^2+dz^2\, .
\end{equation}
It is easy to see that any observer who is at rest with respect to the coordinates  
$r,z,\varphi$ in (\ref{s1.11}) is
a Rindler observer with the acceleration  $w=1/r$. We call
(\ref{s1.11}) the {\it cylindrical Rindler coordinates}, the corresponding
observers are called the {\it cylindrical Rindler observers} to distinguish them
from standard (planar) Rindler observers.
The coordinates (\ref{s1.11}) have a future event horizon which
is topologically ${\cal C}\times R^1$, where $\cal C$ is a light cone
 with the apex at $t=-a$ that  crosses
$t=0$ surface at a circle $x^2+y^2=a^2$. Signals sent from inside of the future event horizon reach
none of the cylindrical Rindler observers.
This follows from the fact that
the Rindler horizon of each particular observer is tangent to the event horizon
in the cylindrical coordinates.
One can also define a past even horizon  for this set of observes
by the reflection with respect to the surface $t=0$.
The bifurcation surface of these horizons, ${\Sigma}$,
is a cylinder.

One can introduce the reduced density matrix $\hat{\rho}$ of the cylindrical Rindler observers by taking the trace over degrees of freedom inside the cylinder $\Sigma$. If 
${\Sigma}$ is considered as an entangling
surface the computations of quantities like $\mbox{Tr}\hat{\rho}_R^n$ require the field  theory to live on
a Euclidean manifold ${\cal M}_n$ which is obtained (in a complete analogy
with the case of planar Rindler coordinates (\ref{s1.5})) by the Wick rotation
in (\ref{s1.11}). This yields
\begin{equation}\label{s1.12}
ds^2=r^2 d\tau^2+dr^2+(a+r \cos \tau)^2d\varphi^2+dz^2\, ,
\end{equation}
where $0 \leq \tau < 2\pi n$ and the condition $0<r \leq b <a$ is implied.
Metric (\ref{s1.12}) coincides with (\ref{1}) and it has a conical singularity
on the cylinder $r=0$ (i.e. on $\Sigma$).

On the other hand, one can also make the coordinate transformation
\be\label{s1.14}
&&x=a+r \cosh \tau \sin \theta \cos\varphi \,  , \, \,
y=a+r \cosh \tau \sin \theta \sin\varphi \, ,\nonumber \\
&&z=a+r \cosh \tau \cos \theta\, ,\, \,
t=r \sinh \tau\, ,
\ee
which changes the metric (\ref{s1.3}) to the form
\begin{equation}\label{s1.15}
ds^2=-r^2 d\tau^2+dr^2+(a+r \cosh \tau)^2(d\theta^2+\sin^2\theta d\varphi^2)\, .
\end{equation}
These coordinates can be called {\it spherical Rindler coordinates} since the
observers which are at rest with respect to (\ref{s1.14}) are the Rindler observes
moving radially toward or out of the center $x=y=z=0$. We call these
observers the {\it spherical Rindler observers}. The future event horizon
of spherical Rindler observers is a three-dimensional cone crossing the sphere $x^2+y^2+z^2=a^2$
at $t=0$ and having the apex at $t=-a$. The past event horizon
is obtained by the reflection  with respect to  the plane $t=0$. The bifurcation surface
of these horizons, ${\Sigma}$,
is a 2-sphere with the position at $t=0$. Studying of the quantum entanglement
across $\Sigma$ with the help of the replica method leads to the Euclidean space (\ref{2}).

In the standard Rindler metric (\ref{s1.5}) the coordinates  belong to the class of Killing frames of reference.
The reduced density matrix then allows a representation
$\hat{\rho}_R\sim \exp(-\hat{H}_R/T_U)$ in terms of a local Rindler Hamiltonian $\hat{H}_R$ which
generates translations along the Killing time $\tau$. Because of this property
one can consider non-integers powers $\hat{\rho}^\alpha_R\sim \exp(-\alpha H_R/T_U)$
which are described by the metric (\ref{s1.8}), where the period of $\tau$ is $2\pi \alpha$.

On the other hand, the metrics (\ref{s1.11}) and (\ref{s1.15}) (and their Euclidean 
counterparts (\ref{1}) and (\ref{2})) are not static.
Therefore, the symmetry under the time translation $t\rightarrow t+ b$ is missing. However, in the Euclidean version of the metric
(\ref{1}) and (\ref{2}) there exists a  symmetry under the global translations, $\tau\rightarrow \tau+2\pi k$, for any integer $k$.
This is all we need in order to construct $\mbox{Tr}\hat{\rho}_R^n$.
As a result of this feature only integer powers of the reduced density
matrices of the cylindrical and spherical Rindler observers allow a geometrical representation. One can
define these matrices in terms of a modular Hamiltonian
$\hat{H}$ as $\rho = \exp(-2\pi \hat{H})$. However, the modular Hamiltonians
for cylindrical and spherical Rindler observers
are essentially non-local operators.
The observation that the metric is not static  is closely related to the fact that the bifurcation
surface $\Sigma$ has the non-vanishing extrinsic curvatures. The latter property is in the main focus of our study here.

\subsection{General definition of squashed cones}\label{gen-d}

Metrics (\ref{s1.11}) and (\ref{s1.15}),  where 
a complete rotation of $\tau$ is $2\pi n$, correspond to orbifolds ${\cal M}_n$ obtained by gluing $n$ copies of the flat space with cuts. The
cuts are made in $t=0$ plane and end on $\Sigma$. Our aim now is to specify a general class of orbifolds ${\cal M}_n$
which we are dealing with. We start with
spaces which appear when the replica procedure is applied to a static spacetime $\cal M$
with the metric
\begin{equation}\label{s1.16}
ds^2=B(x)dt^2+h_{ab}(x)dx^a dx^b~~,
\end{equation}
where $a,b=1,..,d-1$. This case is most interesting from the point of view of its applications.
From now on we consider  the Euclidean signature manifolds, so $B(x)>0$. We choose an entangling surface
$\Sigma$ in a constant time section $\cal H$. The surface $\Sigma$ is a co-dimension  two surface which, in the case of a static spacetime,
has only one  non-vanishing extrinsic curvature. The extrinsic curvature
for a normal vector directed along the Killing vector $\partial_t$ is zero.

We suppose that $\Sigma$ divides a constant $t$ hypersurface $\cal H$
on two parts, say, ${\cal H}_L$ and ${\cal H}_R$, as in the case of the Rindler spacetimes.  The construction of ${\cal M}_n$ requires, first, the 
preparation of a single replica by cutting $\cal M$ along ${\cal H}_L$ (or ${\cal H}_R$). Then $n$ replicas are glued
together along the cuts as was described above.

We would like to understand the properties of the metric of ${\cal M}_n$ near $\Sigma$
in a suitably chosen coordinates and then generalize these properties beyond the class
of static spacetimes.
Let us consider in $\cal H$ the normal Riemann coordinates
$r,y^i$ ($i=1,..,d-2$) with the origin on $\Sigma$. One has
\begin{equation}\label{s1.17}
h_{ab}(x)dx^a dx^b=d\varrho^2+(\gamma_{ij}(y)+2\varrho k_{ij}(y)+O(\varrho^2))dy^idy^j\, ,
\end{equation}
Coordinate $\varrho$ is the geodesic distance from a point on the hypersurface $\cal H$ to $\Sigma$.  The element $\gamma_{ij}(y)dy^idy^j$ is a metric on 
$\Sigma$.
It is easy to show that $k_{ij}(y)$ is the extrinsic
curvature tensor of $\Sigma$  for the unit normal vector $n_a=\delta^r_a$.
($k_{ij}(y)$ is also an extrinsic curvature of $\Sigma$ in $\cal M$.)
It is convenient to introduce a coordinate $\zeta=\sqrt{B}t$,
\begin{equation}\label{s1.18}
d\zeta=\sqrt{B}dt+\zeta w_adx^a\, ,
\end{equation}
where $w_a=\frac 12\partial_aB/B$ are the components of the acceleration vector
of the coordinate frame.
Up to terms of the second order in $\varrho$ and $\zeta$
metric (\ref{s1.16}) near $\Sigma$ takes the form
\begin{equation}\label{s1.19}
ds^2\simeq d\zeta^2+d\varrho^2+(\gamma_{ij}(y)+2\varrho k_{ij}(y))dy^idy^j-2\zeta w_\varrho(y)d\zeta d\varrho
-2\zeta d\zeta w_i(y)dy^i\, .
\end{equation}
This asymptotic behavior depends on the choice of coordinates. One can make an  additional
coordinate transformation
\begin{equation}\label{s1.20}
v^i=y^i-\frac 12 \zeta^2 w^i(y)~~,~~~
\bar{\varrho}=\varrho-\frac 12 \zeta^2 w_\varrho(y)
\end{equation}
to bring (\ref{s1.19}) to a simpler form
\begin{equation}\label{s1.21}
ds^2\simeq dx_1^2+dx_2^2+(\gamma_{ij}(v)+2x_2\, k_{ij}(v))dv^idv^j\, ,
\end{equation}
where we introduced $x_1=\zeta$ and $x_2=\bar{\varrho}$ and omitted the terms
 proportional to $\zeta^2, \bar{\varrho}^2, \bar{\varrho}\zeta$. The next coordinate transformation
\begin{equation}\label{s1.22}
x_1=r \sin \tau\, , \, \,
x_2=r \cos \tau
\end{equation}
brings (\ref{s1.21}) to the form
\begin{equation}\label{s1.23}
ds^2\simeq r^2d\tau^2+dr^2+(\gamma_{ij}(v)+2r \cos\tau k_{ij}(v))dv^idv^j\, .
\end{equation}
One can compare it with (\ref{1}), (\ref{2}) to see that these metrics
 are particular cases of (\ref{s1.23}).

In the case of a static spacetime  the entangling surface
$\Sigma$ has a single non-vanishing extrinsic curvature.
The generalization of (\ref{s1.21}) to surfaces with two non-trivial curvatures
is straightforward:
\begin{equation}\label{s1.24}
ds^2\simeq dx_1^2+dx_2^2+(\gamma_{ij}(v)+2x_p\, k^{(p)}_{ij}(v))dv^idv^j\, ,
\end{equation}
where $p=1,2$, $k^{(p)}_{ij}$ are extrinsic curvatures of $\Sigma$ for
normals $n_p$ ($(n_p)_\mu=\delta_\mu^p$).  After the coordinate transformation (\ref{s1.22}) to the polar coordinate system
this metric becomes
\be
ds^2\simeq r^2d\tau^2+dr^2+\left(\gamma_{ij}(v)+2r \cos\tau k^{(1)}_{ij}(v)+2r \sin\tau k^{(2)}_{ij}(v)\right)dv^idv^j\, ,
\lb{metric}
\ee
which is a generalization of (\ref{s1.23}). The metrics of the type (\ref{metric}) were recently considered in \cite{Lewkowycz:2013nqa}
with a similar motivation to generalize the conical singularity method to the metrics which are not static.

\section{Squashed cones, regularization and curvature invariants}
\setcounter{equation}0

\subsection{Regularization of symmetric cones}

Let us start with a brief review of the regularization method used in \cite{Fursaev:1995ef} in order to calculate the
curvature invariants on a manifold with conical singularities with the rotational 
symmetry. Consider a static Euclidean metric of the following type:
\be
&&ds^2=g(r)d\tau^2+dr^2+\gamma_{ij}(r,v)dv^idv^j\, , \nonumber \\
&&g(r)=r^2+O(r^4)\, , \ \ \gamma_{ij}(r,v)=\gamma_{ij}(v)+O(r^2)\, .
\lb{2.1}
\ee
By setting the period  $2\pi n$ for the angular coordinate $\tau $ 
we get an orbifold  ${\cal M}_n$ with conical singularities at $r=0$, the singular surface $\Sigma$ is 
equipped with intrinsic coordinates $\{v^i\}$ and
intrinsic metric $\gamma_{ij}(v)$. ${\cal M}_n$ can be considered as a limit in the sequence of regular manifolds $\tilde{\cal M}_n$ parametrized by a
regularization parameter $b$.
The family of regular metrics can be obtained by replacing in (\ref{2.1})  the
$g_{rr}$ component to
$\tilde{g}_{rr}=f_n (r,b)$, where the regularization function takes the form
\be
f_n (r,b)=\frac{r^2+b^2n^2}{r^2+b^2}\, .
\lb{2.2}
\ee
Metric (\ref{2.1}) is then recovered in the limit when $b\rightarrow 0$. The limit however should be taken in certain order. Consider in $\tilde{\cal M}_n$ a 
domain $\Omega (r_0)$ around $\Sigma$,  defined as $0<r<r_0$, and change the radial variable, $r=bx$. In terms of the new 
variable $x$ the regularization function becomes
independent of the parameter $b$, $f_n(x)=\frac{x^2+n^2}{x^2+1}$. During
the limiting procedure $f_n(x)$ is fixed but the integral curvature invariants depend on $b$.
This allows one to consider asymptotic series with respect to the parameter $b$.
The integral over the given domain of an invariant $\cal R$ which is
a $m$ order a polynomial of  components of the  Riemann tensor behaves, 
when $b\rightarrow 0$, as 
\be
&&\int_{\Omega(r_0)}\sqrt{g}d^dx ~{\cal R}=\int_0^{2\pi n}d\tau \int_0^{r_0/b}dx 
\sqrt{f_n(x)}\int_\Sigma \sqrt{\gamma}d^{d-2}v \, {\cal R} \nonumber \\
&& =\frac{A_k}{b^k}+\frac{A_{k-1}}{b^{k-1}}+..+A_0 +O(b)\, .
\lb{2.3}
\ee
In the first line in (\ref{2.3}) the integration over $x$ in this limit 
can be extended to infinity. The highest power $k$ of the singular terms 
in these asymptotic series depends on the 
order $m$ of the polynomial $\cal R$. Since spacetime is static integrals  (\ref{2.3}) 
can be considered at arbitrary values of $n$. If one takes the limit 
$n\to 1$, the terms $A_k$ with $k>0$ can be shown to be $O((1-n)^2)$,
while $A_0$  is proportional
$(1-n)$. In applications we are only interested in  the $A_0$ terms. Since the size of the domain $r_0$ can be made arbitrary small, $A_0$ represents a contribution of the conical singularities  and 
is determined by local (intrinsic and extrinsic) geometry on $\Sigma$.

\subsection{Regularization of squashed cones}

The same regularization procedure can be applied to metrics (\ref{1}) and (\ref{2}).
However, if one only replaces the component  $g_{rr}$ with the regularization 
function $f_n(r,b)$, there appears a curvature singularity at $r=0$ in the regularized metric. It can be seen, for instance, from the behavior of the Ricci scalar,  
which diverges as $1/r$. This feature is a manifestation of the 
fact that the
squashed conical singularities are not a direct product of a two-dimensional cone and  $\Sigma$. These geometries near $\Sigma$ have a structure of a warped product of 
the two spaces.  In order to overcome 
the difficulty we have to introduce an additional regularization parameter $p$ in the metric. 
This can be done by changing the power of $r$ in the warped 
factor $(a+r\cos \tau)$ to
$(a+r^pc^{1-p}\cos\tau)$, where $c$ is an irrelevant constant. The singularity at $r=0$ in the regularized metric goes away if $p>2$.
In the present paper we apply the regularization method to integrals quadratic in 
curvatures. Such integrals are regular if $p>1$.

The option we choose is not to introduce $p$ as an independent parameter. It is enough
to assume that it is a function $p(n)$ of $n$, such that $p(n)=n+O(n-1)^2$.
 The simplest choice which we use for the further computations is $p(n)=n$.
Thus, regularized metrics (\ref{1}), (\ref{2}) look as
\begin{equation}\label{1-b}
ds^2=r^2 d\tau^2+f_n (r,b) dr^2+(a+r^n c^{1-n}\cos \tau)^2d\varphi^2+dz^2\, ,
\end{equation}
\begin{equation}\label{2-b}
ds^2=r^2 d\tau^2+f_n (r,b)dr^2+(a+r^n c^{1-n}\cos \tau)^2(d\theta^2+\sin^2\theta d\varphi^2)\, .
\end{equation}
Note that $c$ will not appear in the final result in the limit $n\rightarrow 1$.
We denote geometries (\ref{1-b}), (\ref{2-b}) as $\tilde{\cal M}_n$.

A note of caution regarding the limit $n$ to $1$ should be added. Since
regularized metrics (\ref{1-b}), (\ref{2-b}) depend explicitly
on $\cos \tau$, they cannot be considered 
at arbitrary non-integer values of $n$. At non-integer $n$ there is a jump in extrinsic
curvatures on the hypersurfaces $\tau=0$ and $\tau=2\pi n$. 
Therefore, our prescription is  to do first the computations for an integer $n$ and then analytically continue  $n$ to $1$.  This is the essence of the replica method, as it is 
used in statistical physics. Since $n\geq 2$ for the orbifolds regularizations (\ref{1-b}), (\ref{2-b}) make 
finite all integrals quadratic in curvatures.

\subsection{Integrals for cylindrical and spherical singular surfaces}\label{s3-3}

With these remarks the calculations are pretty  straightforward although a bit tedious.
For the integral of the Ricci scalar we do reproduce the known formula (\ref{0}) earlier established in the case of $O(2)$-symmetric conical singularities.
Below we summarize the results of the calculation for the quadratic combinations of curvature
for regularized metrics (\ref{1-b}), (\ref{2-b}).
When the singular surface is a cylinder (metric (\ref{1-b})) we obtain in the limits
described above
\be
&&\int_{\tilde{\cal M}_n}\sqrt{g}d^4x~R^2\to O(n-1)^2\, \nonumber \\
&&\int_{\tilde{\cal M}_n}\sqrt{g}d^4x~R_{\mu\nu}R^{\mu\nu} \to 4\pi^2\frac{L}{a}(n-1)+O(n-1)^2\, ,\nonumber \\
&&\int_{\tilde{\cal M}_n}\sqrt{g}d^4x~R_{\mu\nu\alpha\beta} R^{\mu\nu\alpha\beta}\to16\pi^2\frac{L}{a}(n-1)+O(n-1)^2\, ,
\lb{2.5}
\ee
where $L$ is the length of a cylinder in the direction $z$. When $\Sigma$ is a sphere (metric (\ref{2-b})) we find 
\be
&&\int_{\tilde{\cal M}_n}\sqrt{g}d^4x~ R^2 \to O(n-1)^2\, \nonumber \\
&&\int_{\tilde{\cal M}_n}\sqrt{g}d^4x~ R_{\mu\nu}R^{\mu\nu}
\to 32\pi^2(n-1)+O(n-1)^2\, ,\nonumber \\
&&\int_{\tilde{\cal M}_n}\sqrt{g}d^4x~ R_{\mu\nu\alpha\beta} R^{\mu\nu\alpha\beta}
\to 64\pi^2 (n-1)+O(n-1)^2\, .
\lb{2.6}
\ee
We note that terms $O(1-n)^2$ contain divergences when $b$ is taken to zero. These terms are not universal and depend on the regularization.

\subsection{A general case of integrals quadratic in curvature}\label{R2}

The above results for a sphere and a cylinder can be used to derive general formulas for the curvatures  for conical singularities on orbifolds. One starts with the case of orbifolds
which are locally flat.
The key observation here is that there exist only two combinations of the extrinsic curvature which can contribute to the surface integrals, $k^2$ 
and
$\tr k^2$. Therefore, the integral of any combination ${\cal R}$ quadratic in Riemann curvature can be expressed as a linear combination of 
integrals of these two
invariants over $\Sigma$
\be
 \int_{\tilde{\cal M}_n}\sqrt{g}dx^4~ {\cal R} \to
 (n-1)\left(\alpha_1 \int_\Sigma \sqrt{\gamma} d^{2}y~ k^2+\alpha_2 \int_\Sigma \sqrt{\gamma} d^2y~ \tr~ k^2\right)+O(n-1)^2\, .
\lb{2.7}
\ee
The unknown coefficients $\alpha_1$ and $\alpha_2$ can be determined by applying (\ref{2.7}) to cases considered in sec. \ref{s3-3}.
When $\Sigma$ is a cylinder we have
\be
\int_\Sigma \sqrt{\gamma} d^2y~ k^2=\int_\Sigma \sqrt{\gamma} d^2y~ \tr ~k^2=2\pi \frac{L}{a}\, ,
\lb{2.8-a}
\ee
while for a sphere
\be
\int_\Sigma \sqrt{\gamma} d^2y~ k^2= 16\pi\, , \ \ \int_\Sigma \sqrt{\gamma} d^2y~ \tr ~k^2=8\pi\, .
\lb{2.8}
\ee
If these results are used together with  (\ref{2.5}) and (\ref{2.6}) we find for the
squashed conical singularities in locally flat geometries the following result: 
\be
&& \int_{\tilde{\cal M}_n}\sqrt{g}d^dx~ R^2\to O(n-1)^2\, , \nonumber \\
&& \int_{\tilde{\cal M}_n}\sqrt{g}d^dx~ R_{\mu\nu}R^{\mu\nu} \to 2\pi (n-1)\int_\Sigma \sqrt{\gamma} d^{d-2}y~k^2+O(n-1)^2\, , \nonumber \\
&& \int_{\tilde{\cal M}_n}\sqrt{g}d^dx~ R_{\mu\nu\alpha\beta}R^{\mu\nu\alpha\beta} \to
8\pi (n-1)\int_\Sigma \sqrt{\gamma} d^{d-2}y~ \tr~ k^2+O(n-1)^2\, .
\lb{2.9}
\ee
This result can be combined with the one obtained earlier in \cite{Fursaev:1995ef} in the case of the symmetric conical singularities, when the extrinsic curvatures are vanishing. We
immediately arrive at formulas (\ref{5.1})-(\ref{5.3})  announced in the Introduction. It should be noted that in (\ref{5.1})-(\ref{5.3})  the cross-terms of the form $\int_\Sigma 
{\cal R}^m k^l$, where $k$ is extrinsic curvature and ${\cal R}$ is the Riemann curvature, are not allowed by dimensional reasons. These terms may appear for the integrals  of cubic and higher combinations of curvature. 

\subsection{Topological and conformal invariants}

In order to demonstrate that our results are robust, and as a consistency check,  we apply obtained relations
to the Euler characteristics and conformal invariants on manifolds
with squashed conical singularities. Let us define
\be\lb{e}
E_4=R_{\mu\nu\alpha\beta}R^{\mu\nu\alpha\beta}-4R_{\mu\nu}R^{\mu\nu}+R^2\, , 
\ee
\be
W^2=R_{\mu\nu\alpha\beta}R^{\mu\nu\alpha\beta}-2R_{\mu\nu}R^{\mu\nu}+\frac{1}{3}R^2\, .
\lb{w}
\ee
On a closed Riemannian 4-dimensional manifold the integral of $E_4$ yields the
Euler characteristics. In four dimensions $W^2$ is the square of the Weyl
tensor. Therefore, the integral of $W^2$ is invariant with respect to the local conformal
transformations of the metric in four dimensions. Since quantities (\ref{e}),(\ref{w}) are 
quadratic polynomials we can define their integrals on ${\cal M}_n$ by applying
(\ref{5.1})-(\ref{5.3}).

The integral of the Euler density over an orbifold with
squashed conical singularities is
\be
\int_{\tilde{\cal M}_n\to {\cal M}_n} \sqrt{g}d^4x~E_4=n\int_{\cal M}\sqrt{g}d^4x~ E_4+8\pi(1-n)\int_\Sigma \sqrt{\gamma}d^2y~{R}_\Sigma\, ,
\lb{3}
\ee
where $R_\Sigma$ is the intrinsic curvature at the 2-surface $\Sigma$. To get (\ref{3})
we used
the Gauss-Codazzi equations which imply that
\be\lb{hf7}
R_\Sigma=R-2R_{ii}+R_{ijij}+k^2-\tr~ k^2\, .
\ee
For closed manifolds Eq. (\ref{3}) can be written as
\be
\chi_4[{\cal M}_n]=n\chi_4[{\cal M}]+(1-n)\chi_2[\Sigma]\, ,
\lb{3a}
\ee
where $\chi_4[{\cal M}_n]$, $\chi_4[{\cal M}]$, $\chi_2[\Sigma]$ are the Euler characteristics of ${\cal M}_n$, ${\cal M}$ and $\Sigma$, respectively.
Equation (\ref{3a}) coincides with the corresponding relation for $O(2)$ symmetric conical 
singularities found in \cite{Fursaev:1995ef}. One can also use (\ref{3a})
for orbifolds which are locally flat. This requires adding corresponding boundary terms
to integral (\ref{3}) which make topological characteristics non-trivial even when
bulk curvatures vanish.

Consider now the integral of the square of the Weyl tensor.  In the limit $n\to 1$ we find
\be
\int_{\tilde{\cal M}_n} \sqrt{g}d^4x~W^2
\to n\int_{\cal M}\sqrt{g}d^4x~W^2+8\pi(1-n)\int_\Sigma \sqrt{\gamma}d^2y~ K_\Sigma+..\, ,
\lb{4}
\ee
where we introduced a conformal invariant
\be
K_\Sigma=R_{ijij}-R_{ii}+\frac{1}{3}R-\left(\tr~ k^2-\frac{1}{2} k^2\right)\, .
\lb{K}
\ee
The conformal invariance of the l.h.s. of (\ref{4}) implies that the term
proportional to $(n-1)$ in the r.h.s is invariant as well (the bulk term in (\ref{4}) is
a conformal invariant). To see the conformal invariance of the terms on $\Sigma$ we 
note that the combination $R_{ijij}-R_{ii}+R/3$ is expressed as a normal projection of the Weyl tensor. One can also show that the integral of $(\tr~ k^2-k^2/2)$ yields another 
conformal invariant.

Both the Euler number and the Weyl tensor squared appear in the logarithmic terms in entanglement entropy of a four-dimensional conformal field theory.
The respective surface contributions (\ref{3}) and (\ref{4})  have been obtained in \cite{Solodukhin:2008dh} by using the arguments 
that involve the holography. Later, for simple geometries in Minkowski spacetime, it was confirmed both by numerical
calculations (for a sphere and a cylinder) \cite{numerical} and by an analytic analysis (for a sphere) \cite{analytic}. We thus confirm this analysis by a direct purely geometric
computation which does not involve any extra assumptions.

\subsection{Integrals in higher dimensions}

In higher dimensions, again by the dimensional reasons, there are still only two possible terms, $\tr k^2$ and $k^2$,  constructed from the extrinsic  
curvature that may contribute to the integrals of quadratic combinations of the Riemann curvature. Thus, the general structure (\ref{2.5}) is valid in a 
higher dimension $d$ although the exact coefficients
may depend on $d$. These coefficients again can be determined by doing the computation for two test surfaces, $S^{d-2}$ and $S^1\times R^{d-3}$. 
We have done these calculations (see details in the Appendix) and we have checked  that in dimensions $d=5$ and $d=6$ the formulas (\ref{2.9}) are not 
changed. We conclude that, likely, the coefficients in (\ref{2.9}) do
not dependent on $d$.  This is similar to what we had in the case of $\Sigma$ with an $O(2)$ symmetry when the terms  are universal. 
The exceptional case is $d=3$  when $\Sigma$ is a curve for which the extrinsic curvature is not defined.

\section{Applications}
\setcounter{equation}0
\subsection{Heat kernel coefficients for squashed cones}
With respect to the reduced density matrix $\hat{\rho}_R$ one can define an effective action $W(n)=-\ln\mbox{Tr}\hat{\rho}_R^n$. For a free field with a wave operator 
$\hD$ 
\be
W(n)=-\frac{1}{2}\int^\infty_{\epsilon^2} \frac{ds}{s}\mbox{Tr}~e^{-s\hD}\, ,
\lb{W}
\ee
where $\epsilon$ is a UV cut-off. The effective action here is defined on an orbifold ${\cal M}_n$ .

The operator $\hD$ is of a Laplace type and one expects that it is well defined on a Hilbert 
space of fields on ${\cal M}_n$ so that the heat trace of 
this operator has a standard asymptotic behavior 
\begin{equation}\label{h.1}
\mbox{Tr}~e^{-s\hD}\sim\frac{1}{(4\pi s)^{d/2}}\sum_{p=0} A_p(\hD)~s^{p}~~
\end{equation}
at small $s$. Here $A_p(\hD)=A_p(n)$ are the heat coefficients which are given by
integrals of local invariant structures on ${\cal M}_n$. 
In the presence of conical singularities the heat coefficient
are represented as
\begin{equation}\label{h.2}
A_p(n)=A_p^{(\mbox{\tiny{reg}})}(n)+A^{(\mbox{\tiny{surf}})}_p(n)\, ,
\end{equation}
where $A_p^{(\mbox{\tiny{reg}})}(n)$ is given by an integral over a regular domain
of ${\cal M}_n$, and $A^{(\mbox{\tiny{surf}})}_p(n)$ is a contribution from the conical singularities.
This contribution is given by an integral over the singular surface $\Sigma$.
The regular part has the same form as in the absence of the conical singularities. Therefore,
$A_p^{(\mbox{\tiny{reg}})}(n)=nA_p(1)$.

By taking into account the structure of $A^{(\mbox{\tiny{surf}})}_p(n)$
in the case of the conical singularities with $O(2)$ isometry we assume that $A^{(\mbox{\tiny{surf}})}_p(n)$
has a simple analytical dependence on $n$ such that one can go to  continuous  values
of $n$ and consider the limit $n\to 1$. Our aim is to determine $A^{(\mbox{\tiny{surf}})}_p(n)$ in the limit
$n\to 1$ when the entangling surface is an arbitrary co-dimension 2 surface.

In order to illustrate our procedure we do it first for the case a scalar field with the operator $\hD=-\Delta+V$, where $V$ is a  potential.
In this case the regular bulk part of the heat kernel coefficients is well known,
\be\label{h.3}
&&A_0^{(\mbox{\tiny{reg}})}(n)=n\, \mbox{vol}({\cal M})\, , \, \, \, A_1^{(\mbox{\tiny{reg}})}(n)=n\int_{{\cal M}}\sqrt{g}d^dx\left(\frac{1}{6}R-V\right)\, , \nonumber \\
&&A_2^{(\mbox{\tiny{reg}})}(n)=n\int_{{\cal M}}\sqrt{g}d^dx\left(\frac{1}{180}R_{\mu\nu\lambda\rho}R^{\mu\nu\lambda\rho}-\frac{1}{180}R_{\mu\nu}R^{\mu\nu}+\frac{1}{2}\left(\frac{1}{6}R-V\right)^2 
\right)\, .
\ee
The surface part of the coefficients was calculated in \cite{Fursaev:1994in} in the case when $\Sigma$ is a fixed point of an Abelian isometry, so that the 
extrinsic curvatures of $\Sigma$ vanish. In  particular, in \cite{Fursaev:1994in} the exact dependence on $n$ of the surface heat kernel coefficients was 
determined.  If we are however interested in the leading in $(1-n)$ behavior of the coefficients then an important observation made in 
\cite{Fursaev:1994ea}  is the following:
\be
 A_p^{(\mbox{\tiny{reg}})}(n)+A^{(\mbox{\tiny{surf}})}_p(n)=A_p^{(\mbox{\tiny{reg}})}
 ({\cal M}_n) +O(1-n)^2\, .
\lb{obs}
\ee
The left hand side contains an exact heat kernel coefficient computed on the conical space while on the right hand side one has the regular, bulk, 
part of the coefficient
extended to a conical space using the formulas (\ref{5.1})-(\ref{5.3}). 
The right hand side should be understood in the sense of the limiting procedure explained above. Ref. \cite{Fursaev:1994ea} dealt with the case of vanishing extrinsic curvatures. 
Here we suggest that this result can be
extended to a general case and determine the surface heat kernel coefficients to leading order in $(1-n)$ using our generalized formulas (\ref{0}) and 
(\ref{5.1})-(\ref{5.3}),
\be
&&A_0^{(\mbox{\tiny{surf}})}(n)=0\, , \, \, A_1^{(\mbox{\tiny{surf}})}(n)=\frac{2\pi}{3} (1-n)\int_\Sigma 1+O(1-n)^2\, , \lb{aa}\\
&&A_2^{(\mbox{\tiny{surf}})}(n)=\frac{2\pi}{3} (1-n)\int_\Sigma\left( \frac{1}{6}R-V+\frac{1}{30}(2R_{ijij}-R_{ii}-2\tr k^2+\frac{1}{2}k^2)\right)+O(1-n)^2\, .\nonumber
\ee
The validity of (\ref{aa}) in the case of $O(2)$ isometry (extrinsic curvature is zero) can be checked by direct comparison of (\ref{aa}) with the exact 
expressions obtained in
\cite{Fursaev:1994in}.
The suggestion about the dependence of the heat kernel coefficients on the extrinsic curvature of the singular surface is our new result. Direct computations 
have to be done to confirm it.

Expressions similar to  (\ref{aa}) for the  $A_2$ coefficient have been also established in the case of the massless Dirac fields and for the gauge invariant combination of the heat coefficients  in the case of gauge fields.
Thus, our suggestion is as well applicable to the heat kernel coefficients of other Laplace type operators on ${\cal M}_n$.

Going back to the scalar fields one can note that in the conformally invariant case, $V=\frac{1}{6}R$, the coefficient $A_2^{(\mbox{\tiny{surf}})}(n)$ is conformally invariant. Below we shall consider the  conformal theories in more detail.

\subsection{Surface terms in conformal anomalies}

In a conformal field theory in $d=4$ the heat kernel coefficient $A_2$, both the bulk and the surface parts, is supposed to be conformal invariant.
By conformal invariance we here mean the invariance under the transformations, $g_{\mu\nu}\rightarrow e^{2\sigma}g_{\mu\nu}$ and $\gamma_{ij}\rightarrow 
e^{2\sigma}\gamma_{ij}$,
of both the bulk metric $g_{\mu\nu}$ and of the induced surface metric $\gamma_{ij}$. In general, for a conformal field theory, the bulk part of the 
coefficient is a linear combination of the Euler density and the Weyl tensor squared
\be
&&A_2=\int_{\cal M} \sqrt{g}d^{4}x(-aE_4+bW^2)\, ,  
\lb{s1}
\ee
where $E_4$ and $W^2$ are defined in (\ref{e}), (\ref{w}).
The coefficients $a$ and $b$ depend on the spin of the field in question. For a real conformal scalar field $a=1/360$ and $b=3/360$. The coefficient (\ref{s1}) 
determines what is called the conformal anomaly.
Under a global rescaling $g\rightarrow \lambda^2 g, \, \gamma\rightarrow \lambda^2\gamma$ the UV finite part of the effective action changes as
\be
W_{\rm fin}(\lambda^2g,\lambda^2 \gamma)=W_{\rm fin}(g,\gamma)+\frac{A_2}{16\pi^2}\ln\lambda\, .
\lb{s11}
\ee
On a conical space the coefficient $A_2$ has a surface part so that the anomaly in (\ref{s11}) contains both the bulk and the surface contributions.
Property  (\ref{aa}) implies that to determine the surface part of $A_2$  we 
can use formulas (\ref{3}) and (\ref{4}). This yields
\be
&&A_2^{(\mbox{\tiny{surf}})}(n)=8\pi(1-n)\int_\Sigma \sqrt{\gamma}d^{2}y
\left(-aR_{\Sigma}+bK_\Sigma \right)+O(1-n)^2\, ,\nonumber \\
&&K_\Sigma=R_{ijij}-R_{ii}+\frac{1}{3}R-(\tr k^2-\frac{1}{2} k^2)\, ,
\lb{s2}
\ee
where $K_\Sigma$ is the conformal invariant.
This expression is our result  for the surface conformal anomaly. It confirms  the form proposed earlier in  \cite{Solodukhin:2008dh} on the basis of 
conformal invariance and the holography. Let us emphasize that this result should hold 
for conformal scalar fields, massless spinor and gauge fields.

\subsection{Non-conformal field theories}

In non-conformal massless   field theories  the heat kernel coefficient $A_2$  is modified by addition of the Ricci scalar squared,
\be
A_2=\int_{\cal M}\sqrt{g}d^{4}x (-aE_4+bW^2+cR^2)\, .
\lb{s3}
\ee
We expect that the surface part of the coefficient should be
\be
A_2^{(\mbox{\tiny{surf}})}(n)=8\pi(1-n)\int_\Sigma \sqrt{\gamma}d^{2}y \left(-aR_{\Sigma}+b K_\Sigma +cR\right)+O(1-n)^2\, ,
\lb{s4}
\ee
where we used our main result (\ref{5.1})-(\ref{5.3}) .
We see that the non-conformally invariant term $R^2$ in the bulk coefficient does not produce any extra extrinsic curvature term in the surface part of the 
coefficient.
In particular, this implies  that in the Ricci flat spacetime the surface heat kernel coefficient is the same as in a conformal field theory.
If the quantum field in question has a mass $m$ then the heat kernel of operator $(\hD+m^2)$ is simply the product $e^{-m^2s}e^{-s\hD}$ so that the mass 
dependence of the
surface coefficients and the surface effective action can be easily restored.  We shall not do this here as it is a trivial exercise.

\subsection{Logarithmic term in entanglement entropy of conformal and non-conformal theories}

Obviously, since for the considered entangling
surfaces the extrinsic curvature is non-vanishing, it is expected to contribute to the entanglement entropy.
This problem was first analyzed in \cite{Solodukhin:2008dh} for the logarithmic terms in the entropy in a conformal field theory in four dimensions.
This analysis is based on the conformal symmetry and the holography and can be summarized as follows.

Entanglement entropy can be computed using the replica method by differentiating the effective action $W(n)$ with respect to $n$,
\be
S=- \mbox{Tr} \hat{\rho}\ln \hat{\rho}= (n\partial_n-1)W(n)|_{n=1}\, .
\lb{s5}
\ee
The UV divergences of the entropy can be deduced from the surface heat kernel coefficients we have just derived. For a quantum field of mass $m$ we find that
\be
S=\frac{N_s\, A(\Sigma)}{48\pi \epsilon^2}+\frac{1}{2\pi} 
\int_\Sigma \sqrt{\gamma}d^{2}y(aR_\Sigma-b K_\Sigma -c R +\frac{1}{12}m^2 D_s)\ln\, \epsilon\, ,
\lb{s6}
\ee
where $N_s$ is the total number of physical degrees of freedom in the theory of spin $s$ while $D_s$ is the dimension of the representation of spin $s$. 
We have included here the mass term since its contribution appears to be universal for fields of any spin.
In the conformal case ($c=0$ and $m=0$) equation  (\ref{s6}) agrees with the result derived in
\cite{Solodukhin:2008dh}. The logarithmic term in (\ref{s6}) for a  non-conformal field theory is our new result. We see from (\ref{s6}) that
in a Ricci flat spacetime and for a massless field the non-conformal term in (\ref{s6}) vanishes and the logarithmic term is exactly the same as in a conformal field theory.
This agrees with the numerical computation of the logarithmic term for a scalar field in Minkowski spacetime made in \cite{numerical}.

\subsection{Holographic formula of entanglement entropy 
for generic AdS gravities}\label{HF}

Entanglement entropy $S(\Sigma)$ in conformal field theories (CFT) which admit a dual description
in terms of an anti-de Sitter gravity allow a remarkable representation 
known as a holographic formula \cite{Ryu:2006bv} 
\be\lb{hf1}
S(\Sigma)=\frac{A(\tilde{\Sigma})}{ 4G_{(d+1)}}\, .
\ee
Here $G$ is the Newton constant in the dual gravity theory. The dual theory has one dimension higher than the conformal field theory.  The holographic formula is defined in terms of 
the volume $A(\tilde{\Sigma})$ of a 
minimal codimension 2 hypersurface $\tilde{\Sigma}$ in the bulk AdS spacetime with the condition that the asymptotic boundary of $\tilde{\Sigma}$ belongs to a conformal
class of $\Sigma$. There are infrared divergences in (\ref{hf1}) which 
according to the AdS/CFT 
dictionary \cite{ads/cft} are related to the ultraviolet divergences of the entanglement entropy  on the CFT side.

Formula (\ref{hf1}) is valid for theories where the dual gravity action 
$I[{\cal M}^{(d+1)}]$ has the Einstein form  with a negative cosmological constant. Our method allows one to make the prediction 
about the holographic formula when the bulk gravity includes terms 
quadratic in curvatures
\be\lb{hf2}
I[{\cal M}^{(d+1)}]=-\int_{{\cal M}^{(d+1)}}\sqrt{g}d^{d+1}x~
\left[\frac{R}{ 16\pi G_{(d+1)}}+2\Lambda+
a~R^2+b~R_{\mu\nu}R^{\mu\nu} +c~ R_{\mu\nu\alpha\beta}R^{\mu\nu\alpha\beta}\right] .
\ee
In what follows we look for a  modification of (\ref{hf1}) for 5d gravity (the dual CFT is 4-dimensional).
We use the arguments of \cite{Fursaev:2006ih} for the origin of the holographic formula.
The basic idea here is that in the replica method the $n$-th power of the reduced CFT 
density matrix $\tr~\hat{\rho}^n$ is determined by an AdS `partition function' which in the semiclassical approximation is just an action on an orbifold  ${\cal M}^{(d+1)}_n$
constructed from $n$ replicas of ${\cal M}^{(d+1)}$. Conical singularities of ${\cal M}^{(d+1)}_n$ are located on a codimension 2 hypersurface $\tilde{\Sigma}$ whose 
boundary is conformal to the entangling surface $\Sigma$. According to this prescription
\be\lb{hf3}
S(\Sigma)=-(n\partial_n-1)\ln Z(n)_{n\to 1}\, ,
\ee
where in the semiclassical approximation
\be\lb{hf4}
-\ln Z(n) \sim 
I[{\cal M}^{(d+1)}_n]\, .
\ee
One can use now results (\ref{5.1})-(\ref{5.3}) for the action on squashed conical singularities to get 
from (\ref{hf3}),(\ref{hf4}) the following formula:
\be\lb{hf5}
S(\Sigma)=\frac{A(\tilde{\Sigma}) }{ 4G_{(d+1)}}+
4\pi \int_{\tilde{\Sigma}} \sqrt{\gamma}d^{d}y~\left[2aR+b 
\left(R_{ii}-\frac{1}{2} k^2\right)+
2c (R_{ijij}-\tr~ k^2)\right]\, .
\ee
which reduces to (\ref{hf1}) when $a=b=c=0$. Subadditivity property of entanglement entropy 
and arguments of \cite{Fursaev:2006ih} require that $\tilde{\Sigma}$ is a hypersurface 
at which
the functional in the r.h.s of (\ref{hf5}) has a minimum. 

We will leave studying consequences of Eq. (\ref{hf5}) for a future work and end this section
with a discussion of the $d=5$ Gauss-Bonnet gravity which is a particular case of (\ref{hf2}).
For this type of the theory the constants $a,b,c$ are related to each other and expressed in terms of a single parameter $\lambda$ as $a=2\lambda$, $b=-4\lambda$, $c=\lambda$. 
Combination of the quadratic terms is just an extension to 5 dimensions of the Euler 
density (\ref{e}).
The holographic formula for the bulk Gauss-Bonnet gravity 
\be\lb{hf6}
S(\Sigma)=\frac{A(\tilde{\Sigma}) }{ 4G_{(5)}}+
8\pi \lambda\int_{\tilde{\Sigma}} \sqrt{\gamma}d^{3}y~\hat{R}\, ,
\ee
where $\hat{R}$ is the scalar curvature of $\tilde{\Sigma}$. To get (\ref{hf6})
from (\ref{hf5}) we used the Gauss-Codazzi equations (\ref{hf7}).
Formula (\ref{hf6}) has been suggested in \cite{Fursaev:2006ih} for surfaces with 
vanishing extrinsic curvatures and for a generic surfaces in the Gauss-Bonnet gravity 
in \cite{Hung:2011xb},\cite{deBoer:2011wk}. It is also known as the 
Jacobson-Myers functional.

The fact that our method reproduces the modified holographic formula for the Gauss-Bonnet gravity
gives a further support to this formula.

\subsection{Classical gravitational entropy}

Usually the classical gravitational entropy is attributed to the surfaces which are the  Killing horizons.
The appropriate metric is supposed to possess a  symmetry generated by a time-like Killing vector $\xi_{(t)}$ so that
this vector is null, $\xi_{(t)}^2=0$ on the surface $\Sigma$ in question. One way to derive the corresponding
gravitational entropy is to analytically continue the metric to the Euclidean signature and allow the Euclidean time $\tau$ to 
have an arbitrary periodicity $\beta$. Since the surface $\Sigma$ is a stationary point of the isometry generated by the Killing vector
for a generic $\beta$ there appears a conical singularity at $\Sigma$. The condition that the singularity is absent fixes the Hawking temperature
$\beta^{-1}_H$ of the horizon while the differentiation of the gravitational action with respect to the angle deficit gives the gravitational entropy.
This procedure works in all known cases and gives correctly the horizon entropy.

The analysis of the present paper allows to generalize this prescription and assign some gravitational entropy with surfaces which are not Killing horizons 
in the usual sense. The spacetime metric then can be time dependent. The only condition which we have to impose is that in the Euclidean signature the spacetime metric should possess  a discrete symmetry $\tau\rightarrow \tau+2\pi n$ for any integer $n$.  Suppose that  a surface $\Sigma$ is a fixed point of this discrete isometry.  
Then we can close the Euclidean time with a period $\beta=2\pi n$, $n$ is integer, and repeat all other steps as in the case of the Killing horizons.
The only new thing is that we have to first do all the calculations for an integer $n$, differentiate with respect to the angle deficit and only then take the limit 
$n\to 1$.  This construction appears to be consistent with the proposal in \cite{Lewkowycz:2013nqa} although it does not require existence of any boundaries,
the entropy comes out as a local property of the surface $\Sigma$.
  
In a gravitational theory described by action (\ref{hf2}) the gravitational entropy associated with a surface $\Sigma$ in the procedure just described
takes exactly the form (\ref{hf5}). We stress that it is a purely classical entropy analogous to the Bekenstein-Hawking entropy in the case of the Killing horizons.
Taking the UV renormalization procedure \cite{Fursaev:1994ea} this classical entropy may be an important   element in the on-going discussion (see \cite{Bianchi:2012ev}, \cite{Myers:2013lva})  of the finite entanglement  entropy associated to generic surfaces.

\section{Conclusions}
\setcounter{equation}0

The aim of our work is to check whether one can define a sensible geometry on manifolds with the generalized conical singularities in a distributional sense. We present a method of how it can be done which is applicable to a general class of singularities. 
 We have checked the consistency of the method for the topological and
 conformal invariants. In these cases  we have found an agreement of the earlier  holographic studies of entanglement entropy 
with the distributional nature of the squashed cones. This gives a further support
to the holographic description of entanglement entropy. One of immediate by-products of 
our analysis is a suggestion made in subsection \ref{HF} of a holographic formula for entanglement entropy  in the theories whose gravity dual has an action which is an arbitrary combination of terms quadratic 
in curvature.

Since the conical singularity method 
is applicable to co-dimension 2 surfaces, which are not necessarily bifurcation surfaces
of event horizons, one may attribute to these surfaces an entanglement entropy, as was suggested in 
\cite{Fursaev:2007sg}, \cite{Balasubramanian:2013rqa}, 
 \cite{Bianchi:2012ev}, \cite{Myers:2013lva}.

It should be stressed that
there is an essential
difference in squashed and symmetric cones. 
For the $O(2)$ symmetrical conical singularities there is a useful factorization formula
suggested in \cite{Fursaev:1995ef} 
for the Riemann tensor
\begin{equation}\label{c1}
^{(n)}R^{\mu\nu}_{~~\lambda\rho}=R^{\mu\nu}_{~~\lambda\rho}+
2\pi(n-1)((n^\mu \cdot n_\lambda)((n^\nu \cdot n_\rho)-(n^\nu \cdot n_\lambda)((n^\mu \cdot n_\rho))\delta_\Sigma+O(n-1)^2\, .
\end{equation}
Here 
$R^{\mu\nu}_{~~\lambda\rho}$ is a regular part of the curvature, $(n^\mu \cdot n_\nu)=\sum_{i} n^\mu_i n^i_\nu$, 
$\delta_\Sigma$ is a covariant delta-function with a support on the singular surface 
$\Sigma$.
As we said, (\ref{c1}) is quite useful in applications. Unfortunately, as one can conclude
from our results this simple factorization is not applicable in the case of the squashed conical singularities.
It also cannot be modified to describe correctly the terms which depend
on the extrinsic curvatures of $\Sigma$.

We carried out the computations for the integrals of  polynomials quadratic in curvatures in different dimensions. Since the suggested method  yields  the self-consistent results we believe it can be applied to
higher powers  of the Riemann curvature in various dimensions. Some work in this direction is in progress.

Let us finish with our comments on some recent papers which appeared while the present work was in the final stages of preparation. Two other new papers appeared
which  overlap with some results reported here. The spherical Rindler 
horizons discussed in Sec. \ref{RH} have been also introduced in \cite{Balasubramanian:2013rqa}. The asymptotic form of the metric (\ref{gen-d})
near a codimension 2 hypersurface with the non-vanishing extrinsic curvatures, see 
Sec. \ref{gen-d}, was also used in \cite{Lewkowycz:2013nqa}. The authors of \cite{Lewkowycz:2013nqa} also suggest an interesting alternative derivation of the holographic formula (\ref{hf1}). They use a conformal transformation from the singular boundary
manifolds which appear in the replica method on the CFT side to the non-singular ones. This 
yields
the gravity partition function with the non-singular boundary conditions and the non-singular 
background geometries in the bulk. It would be interesting to find the verifiable  predictions of 
\cite{Lewkowycz:2013nqa} for the bulk gravity theories in the form (\ref{hf2}) and compare them 
with (\ref{hf5}) derived in this paper. This question has been analyzed in \cite{Chen:2013qma},\cite{Bhattacharyya:2013jma}. Results of \cite{Bhattacharyya:2013jma} show that extension of
\cite{Lewkowycz:2013nqa} to higher derivative gravities requires small
extrinsic curvatures. The method presented here does not have restrictions on the
curvatures.

\bigskip
\section*{Acknowledgements}
The work of D.F. and A.P. is supported in part by RFBR grant 13-02-00950. S.S. would like to thank Thibault Damour
for hospitality at IHES, Bures-sur-Yvette, during  the final stage of this project. 
A.P. acknowledges the support from the Heisenberg-Landau Programm. The authors are grateful to Aitor Lewkowycz, Rob Myers, and Aninda Sinha for useful communications.

\newpage
\appendix

\section{Integrals in 5 and 6 dimensions}
\setcounter{equation}0

Here we present calculations of the regularized
integrals which correspond to the case when ${\cal M}_n$ is obtained 
from $n$ copies of the Minkowski spacetime. The copies are glued along cuts
which meet on a codimension 2 hypersurface $\Sigma$. Thus, $\Sigma$ is a singular surfrace where conical singularities are located. We present 
results for different choices of $\Sigma$ in five and six dimensions. The regularized
geometry is denoted as $\tilde{\cal M}_n$. It is convenient to introduce the following notations:
\be\lb{a1}
I_1=\int_{\tilde{\cal M}_n}\sqrt{g}d^dx~R^2\, ,~~
I_2=\int_{\tilde{\cal M}_n}\sqrt{g}d^dx~R_{\mu\nu}R^{\mu\nu}\, ,~~
I_3=\int_{\tilde{\cal M}_n}\sqrt{g}d^dx~R_{\mu\nu\alpha\beta}R^{\mu\nu\alpha\beta}\, .
\ee
\be\lb{a2}
K_1=\int_\Sigma \sqrt{\gamma}d^{d-2}y~ k^2\, ,~~
K_2=\int_\Sigma \sqrt{\gamma}d^{d-2}y~ \tr~ k^2\, .
\ee
The dimensionality $d$ is 5 or 6. In all the cases given below relations (\ref{2.9}) are satisfied.

\bigskip

{\bf A1. $d=5$}.  Let $\Sigma$ be a hypersphere $\Sigma=S^3$ of the radius $a$, then 
\be
&&I_1\to O(n-1)^2\, \nonumber \\
&&I_2\to
36\pi^3 a(n-1)+O(n-1)^2\, ,\nonumber \\
&&I_3\to 48\pi^3 a(n-1)+O(n-1)^2\, ,
\ee
\be
K_1=18\pi^2 a \, ,~~K_2=6\pi^2 a ~\, .
\ee

\bigskip

{\bf A2. $d=5$}.
Let $\Sigma$ be a hypercylinder $\Sigma=S^2 \times R^1$ of the radius $a$ and the length 
$L$, then
\be
&&I_1\to O(n-1)^2\, ,\nonumber \\
&&I_2\to 32\pi^2 L(n-1)+O(n-1)^2\, ,\nonumber \\
&&I_3\to 64\pi^2 L(n-1)+O(n-1)^2\, ,
\ee
\be
K_1=16\pi L ~~,~~K_2=8\pi L ~\, ,
\ee
Note that the radius $a$ of $S^2$ does not appear directly in these formulas.

\bigskip

{\bf A3. $d=5$}.
For $\Sigma=S^1\times R^2$, where $S^1$ has the radius $a$, and $R^2$ is a square of the
area $L^2$ one finds
 \be
&&I_1\to O(n-1)^2\, ,\nonumber \\
&&I_2\to 4\pi^2 \frac{L^2}{a}(n-1)+O(n-1)^2\, ,\nonumber \\
&&I_3\to 16\pi^2 \frac{L^2}{a}(n-1)+O(n-1)^2\, ,
\ee
\be
K_1=2\pi \frac{L^2}{a} ~~,~~K_2=2\pi \frac{L^2}{a} ~\, .
\ee

\bigskip

{\bf A4. $d=6$}.
If $\Sigma$ is a hypersphere $\Sigma=S^4$ of radius $a$
\be
&&I_1\to O(n-1)^2\, \nonumber \\
&&I_2\to \frac{256}{3}\pi^3 a^2(n-1)+O(n-1)^2\, ,\nonumber \\
&&I_3\to \frac{256}{3}a^2\pi^3(n-1)+O(n-1)^2\, ,
\lb{2s4}
\ee
\be
K_1=\frac{128}{3}\pi^2 a^2~~,~~K_2=\frac{32}{3}\pi^2 a^2~\, .
\ee

\bigskip

{\bf A5. $d=6$}.
If $\Sigma$ is a hypercylinder $\Sigma=S^3\times R^1$ of the radius $a$ and the length 
$L$
 \be
&&I_1\to O(n-1)^2\, ,\nonumber \\
&&I_2\to 36\pi^3 a L(n-1)+O(n-1)^2\, ,\nonumber \\
&&I_3\to 48\pi^3 a L(n-1)+O(n-1)^2\, ,
\lb{2s3}
\ee
\be
K_1=18\pi^2 a L~~,~~K_2=6\pi^2 a L~\, .
\ee

\bigskip

{\bf A6. $d=6$}.
If $\Sigma$ is a product $\Sigma=S^2\times R^2$ of a 2-sphere of radius $a$ and a square 
of length $L$
 \be
&&I_1\to O(n-1)^2\, ,\nonumber \\
&&I_2\to 32\pi^2  L^2(n-1)+O(n-1)^2\, ,\nonumber \\
&&I_3\to 64 \pi^3 L^2(n-1)+O(n-1)^2\, ,
\lb{2s2}
\ee
\be
K_1=16\pi L^2 ~~,~~K_2=8\pi L^2~\, .
\ee
Note that the radius $a$ of $S^2$ does not appear directly in the formulas.

\bigskip

{\bf A7. $d=6$}.
For $\Sigma$ being a product $\Sigma=S^1\times R^3$ of a circle of radius $a$ and a cube 
of length $L$ one finds
 \be
&&I_1\to O(n-1)^2\, \nonumber \\
&&I_2\to 4\pi^2  \frac{L^3}{a} (n-1)+O(n-1)^2\, ,\nonumber \\
&&I_3\to 16 \pi^2 \frac{L^3}{a}(n-1)+O(n-1)^2\, ,
\lb{2s1}
\ee
\be
K_1=2\pi \frac{L^3}{a}~~,~~K_2=2\pi \frac{L^3}{a}~\, .
\ee

\end{document}